# Statistical Traffic State Analysis in Large-scale Transportation Networks Using Locality-Preserving Non-negative Matrix Factorization

Yufei HAN and Fabien Moutarde

CAOR, MINES-ParisTech, 60 Boulevard Saint-Michel, 75006, Paris

Yufei.Han@mines-paristech.fr, Fabien.Moutarde@mines-paristech.fr

## Abstract

Statistical traffic data analysis is a hot topic in traffic management and control. In this field, current research progresses focus on analyzing traffic flows of individual links or local regions in a transportation network. Less attention are paid to the global view of traffic states over the entire network, which is important for modeling large-scale traffic scenes. Our aim is precisely to propose a new methodology for extracting spatio-temporal traffic patterns, ultimately for modeling large-scale traffic dynamics, and long-term traffic forecasting. We attack this issue by utilizing Locality-Preserving Non-negative Matrix Factorization (LPNMF) to derive low-dimensional representation of network-level traffic states. Clustering is performed on the compact LPNMF projections to unveil typical spatial patterns and temporal dynamics of network-level traffic states. We have tested the proposed method on simulated traffic data generated for a large-scale road network, and reported experimental results validate the ability of our approach for extracting meaningful large-scale space-time traffic patterns. Furthermore, the derived clustering results provide an intuitive understanding of spatial-temporal characteristics of traffic flows in the large-scale network, and a basis for potential long-term forecasting.

# 1. Introduction

Most traffic information systems make use of floating-car data collected from distributed probing vehicles [1][2][3] as a major data feed for quantitative traffic state evaluation. Acquired floating-car data are aggregated over small time periods (5-15 mn) to estimate traveling time of vehicles, in order to identify traffic states (congestion or free-flowing) for each link. For urban transportation network of decent scale, the traffic management department processes real-time traffic information from thousands of links simultaneously, which is an overwhelming task. Therefore, automatic analysis of traffic information, e.g. unveiling characteristics of traffic flow variations [4][5][6][7][8] is necessary for efficient management strategies and adjusting demands of traffic sources.

Most published works on traffic data analysis focus on modeling temporal dynamics for individual links (either in arterial networks or highways) using model-driven [9][10][11][12][13] or data-driven methods [14][15][16][17][18][19][20][21][22][23][24]. The model-driven methods, like Cellular Automata [12] and other underlying physical models [9][10][11][13], are usually equipped with parameters that are calibrated with structural assumptions to simulate temporal evolution of traffic states. Excellent and as they are for modeling free way or arterial links, the model-driven methods present less efficiency in modeling urban traffics. The velocity flow field of urban transportation is easily subject to the fluctuations induced by intersections of links, traffic signals at the crossings and so on. These fluctuations lead to spatio-temporal traffic events. It is thus difficult to find a local stationary regime for the velocity using the physical models. In contrast, data driven approaches describe statistical dependencies using the block-box machine learning methodologies. They are more popular due to relaxation of prior assumptions during modeling traffic dynamics. Kalman filter [25] and ARMA (Autoregressive Moving Average) [26], originated from state space theory, are used to predict temporal variations of traffic flows [14][15][16][17][18]. It is an extension of these linear models from sequential signal processing to traffic domain. Efficient as they are for short-term temporal prediction, they can not track nonlinear fluctuations of traffic flows. In [19][20][21][22], neural networks [27] and hybrid non-linear dynamic systems are used to approximate short-term non-linear variations of traffic states. Due to the intrinsic multiple-input and multiple-output (MIMO) structures, neural networks can integrate spatial-temporal correlations between local links into a computational framework. In [23][24], spatial correlations between local links are considered in Markov Random Field [28] and Multi-Agent System [29] based traffic models. These inspiring works concatenate global structural information of transportation networks to improve descriptive power of traffic flow models.

Extending both the model-driven and data-driven methods to large-scale urban network, we need to tackle curse-of-dimensionality caused by enlarged scale of the modeling target. Increasingly more parameters are needed to capture details about temporal dynamics of thousands of links, which increases intrinsic complexity of the built traffic model. Therefore, it is necessary to introduce regularization terms, providing prior knowledge about global configuration of traffic states over the entire network and serve as consistency constraints of the spatio-temporal congestion structure, e.g. co-occurrence of congestion in the network during specific time intervals. Furthermore, local regions with independent traffic flow behaviors can be treated separately in a divide-and-conquer manner. Therefore, mining the spatial configuration patterns of congestion and large-scale macroscopic traffic dynamics over the entire network is highly informative for constructing the computationally tractable models to describe local fringes of large-scale traffic scenes. Such macroscopic view of traffic flow configurations can be also used to identify bottleneck of transportation networks, in order to improve traffic management strategies. Besides, drivers can make use of the global traffic state information to optimize their traveling plan ever before they leave their own garage. Nevertheless, little progress has been reported on analyzing global congestion configurations of large networks. We attack this issue by performing clustering analysis of traffic configurations over the entire large-scale network simultaneously. We define the network-level traffic state as a multi-dimensional vector containing traffic states of all local links. A matrix factorization based dimension reduction method named as Locality-Preserving Non-negative Matrix Factorization (LPNMF) [39], is adopted to derive a compact representation of high-dimensional network-level traffic states. K-means clustering [48] performed on the derived compact LPNMF representation provides intuitive understanding of typical spatial configuration patterns of global traffic states and large-scale traffic dynamics of network-level traffic states contained in the data. The flowchart of this work is illustrated in Figure 1.

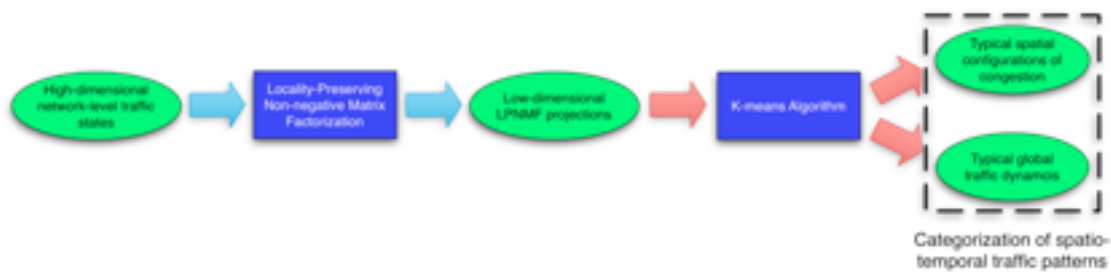

*Figure 1.* *The flow chart of the proposed clustering methodology*

This article is organized as follows. Section 2 introduces LPNMF employed in the analysis. Section 3 presents the simulated traffic data of a large-scale urban network, used as data source in the following analysis. In Section 4, we illustrate detailed clustering results of spatial configuration patterns obtained through the LPNMF projection. Based on the LPNMF based representation, Section 5 further performs a clustering analysis on temporal behaviors of network-level traffic states. Section 6 draws some conclusions and discusses our future work.

## 2. Traffic data mining with Locality-Preserving Non-negative Matrix Factorization

## 2.1. Basic scheme of Non-negative Matrix Factorization (NMF)

LPNMF is an extension of basic NMF [30][31][32][33] by introducing constraints on topological structures of the derived projection subspace. In this section, we introduce basic principles of extracting a flexible representation from original network-level traffic states using the NMF-like scheme. NMF is a particular matrix factorization algorithm, which is in the same family of techniques as the well-known PCA (Principle Component Analysis) [34]. As mentioned in Section 1, *i-th* entry of a network-level traffic state corresponds to the traffic state on *i-th* link of the network. In urban traffic sequences, the number of links in any network of decent scale is often over one thousand. Thus, the network-level traffic state has a rather dense data structure. Assuming *m* samples of n-dimensional network-level states are stored into the column space of a $n \times m$ matrix **X**, NMF factorizes **X** as a product of a $n \times s$ non-negative loading matrix **M** and a $s \times m$ non-negative scoring matrix **V**, in order to minimize the Frobenius norm [35] of the reconstruction error between **X** and the product of **M** and **V**:

$$(M,V) = \underset{M \geq 0, V \geq 0}{\arg\min} \|X - MV\|_F \quad (1)$$

Each column of **V** is the NMF projection of the corresponding network-level traffic states. *s* is the dimensionality of the NMF projection subspace that is spanned by columns of **M**. Normally, *s* is set to be much less than the row length of **X**, thus **V** forms a low-dimensional representation of network-level traffic states. The specificity of NMF is the non-negativity constraint on **M** and **V**. Each network-level traffic state $\mathbf{X_j} \in \mathbf{R^n}$ is approximated by an additive linear superposition of the column space of **M** in NMF [31][32] as in Eq.2:

$$X_j = \sum_{i=1}^{s} M_i V_{i,j} \quad (2)$$

where $\mathbf{X_j}$ and $\mathbf{M_i}$ is the *j-th* column of **X** and the *i-th* column of **M** respectively. $V_{i,j}$ is the element located at the *j-th* column and *i-th* row of **V**. Treating columns of **M** as the learned base for reconstructing the network-level traffic states, they represent typical structural patterns of traffic configurations. $V_{i,j}$ represents to which degree the *j-th* network-level traffic state vector is associated with the spatial configuration pattern of local traffic states represented by $\mathbf{M_i}$. For example, if $\mathbf{M_i}$ can better represent the *j-th* network-level traffic state, $V_{i,j}$ will take the largest value in the *j-th* column of **V** [36]. The row-wise average $\frac{1}{m}\sum_{j=1}^{m}V_{i,j}$ evaluates the importance of the corresponding NMF basis vectors $\mathbf{M_i}$ in representing the spatial congestion configuration. In this sense, the additive combination shown in Eq.2 leads to a part-based decomposition of the network-level traffic states. Localized groups of entries in each basis vector $\mathbf{M_i}$ with distinctively large magnitudes indicate typical patterns or important components of the original data representation. Benefited from the property, NMF is usually used for extracting semantic components of objects from images [30][32] and latent topics from text documents [36][39]. Motivated by the sounding properties of NMF, we use it rather than PCA to investigate spatial patterns and dynamic properties of network-level traffic states.

An iterative procedure, named as MU (Multiplicative Update) [30][33], is used to solve NMF optimization. In each iteration of MU, either of **M** or **V** is fixed alternatively, the other one is then updated by solving a non-negativity constrained least square problem based on KKT theorem [30][31]. Given the dimensionality of NMF projection as *s*, each iteration of MU has a computation cost in $O(nsm)$. As reported in [30], MU converges to the optimum solution with definite iterations. In our work, MU generally takes 600 iterations before its convergence, much less than the number of samples contained in the data matrix. Therefore, MU has better computational efficiency than SVD in our case. With finer tuning of matrix multiplication using Strassen's algorithm or Coppersmith-Winograd approach [30] [33], computational efficiency of MU can be improved in a further step.

## 2.2. Locality-Preserving Penalty for Non-negative Matrix Factorization

For clustering analysis, we want that geometrical structure of the projection space is consistent with the intrinsic characteristics of the traffic data. In particular, we need that projections of network-level traffic states are close to each other, if they are similar in the original high-dimensional space. The consistence of geometrical structures (distance measures between data points) is of utter importance for clustering analysis and dynamic modeling [37][38]. Any artifacts introduced into distance measures in the projection space could change cluster assignments or temporal dynamic patterns. Motivated by this idea, we propose to use a regularized NMF [39][40] to derive the representation of network-level traffic states, named as LPNMF. It aims to minimize the following objective function $O$, as shown in Eq.3:

$$O = \|X - MV\|_F^2 + \lambda Tr(VLV^T) \tag{3}$$

$$L = D - W \tag{4}$$

$$D_{i,i} = \sum_j W_{i,j} \tag{5}$$

where $Tr$ is the trace of a matrix and $\lambda$ is the regularization parameter. The first term is the Frobenius reconstruction error as illustrated in Eq.1, while the second one is the structural regularization of NMF projections. In this term, **L** is called Graph Laplacian [41][42] as defined in Eq.4. In the matrix **W**, the element $W_{i,j}$ located at *i-th* row and *j-th* column, is the pair-wise similarity measure matrix between the *i-th* and *j-th* network-level traffic state vectors, corresponding to the *i-th* and *j-th* column of **X**. According to Eq.5, **D** is a diagonal matrix whose entries are column sums of **W**. Graph Laplacian originates from spectral graph theory [43][44][45]. By adding the Graph Laplacian based constraints, the obtained low-dimensional representation **V** is calibrated to have similar geometrical structures as the original data **X** without increasing further computation cost. Based on this property, we can unveil the characteristics of global traffic states more efficiently in the low-dimensional manifold **V** without loss of intrinsic data distribution information.

### 2.3. Distance measure between network-level traffic states

To perform LPNMF, we need to define a similarity measure between network-level traffic states that evaluates differences between spatial configurations of local traffic states. The traffic state of one link is usually closely correlated with its up-stream or downstream nearest neighbors with the same orientation of traffic flows. For example, the links $u_i^j$ and $d_i^m$ are upstream and down-steam nearest neighbors of the link *i* respectively. Assuming the link *i* fell into heavy traffic congestion, the links $u_i^j$ and $d_i^m$ are more likely to be congested than those far from the link *i*. Motivated by the property, we propose a weighted fusion scheme among traffic states of geometrical neighborhoods to derive the similarity measure. We firstly calculate link-wise differences of traffic states between corresponding links. For each link *i*, we then obtain a weighted sum of the link-wise difference values with respect to the link *i* and its neighbors, which is defined to be local variation $v_i$ of traffic states around the current link, as denoted in Eq.6:

$$v_i = \sum_j w_j^u a(u_i^j) + \sum_m w_m^d a(d_i^m) + w^i a(i) \tag{6}$$

*a* is the link-wise difference of traffic states between the corresponding link. $w_j^u$, $w_m^d$ and $w^i$ are the weights respectively attached to up-stream neighbors, downstream neighbors and the current link *i*. After that, we map $\{v_i\}$ into [0,1] using a Gaussian kernel in Eq.7 as the similarity measure between two network-level traffic states:

$$S = e^{-\frac{\sum_i v_i}{2\delta^2}} \tag{7}$$

To normalize range of the weighted sum, the sum of all weights is required to be 1.

The weight $w^i$ corresponding to the link *i* should be the largest one. Weights of the neighboring links can be designed to be proportional to degrees of traffic state correlation between one specific neighboring link and the current link *i*. In this article, all neighboring links are evaluated with the same weight value. By performing weighted fusion of local neighborhoods in the network, the derived similarity measure not only represents traffic state variations between corresponding links but also indicates the spatial correlations between local neighborhoods. We feed this similarity measure into the matrix **W** in Eq.4 and derive the regularized low-dimensional representation of the network-level traffic states.

## 3. Metropolis simulation and IAURIF database

### 3.1 Metropolis traffic simulation software

The benchmark IAURIF database used to verify the validity of the proposed LPNMF based method is generated by simulating real-traffic sequences of a large-scale traffic network using Metropolis [46][47]. Metropolis is a planning software designed to model urban transportation systems. It allows the user to study impacts of transportation management policies for metropolitan areas and their fringes in a time-dependent framework. Metropolis simulates commuters' traveling behaviors and congestion in urban areas. The core of the simulation system is a dynamic simulator that integrates joint commuters' departure time and their choices of routes in the transportation network [46]. During simulation of traffic sequences, each commuter is characterized by specific parameter values individually [46][47]. At any moment, locations of all commuters are known. Given traveling plans in Origin-Destination (O-D) matrix, commuters choose the shortest path dynamically from their current locations to destination. Traveling time of each commuter on a specific link is estimated based on queuing theory, by minimizing a cost function that achieves trade-off between arriving close to the desired arrival time to incur congestion and arriving early/late compared with the desired arrival time to avoid congestion [46]. Congestion in the network is modeled at a macroscopic level. The congestion laws deciding travel delays of each link depend on the setting of incoming traffic flow during a time period and average rate of occupancy [46][47]. To launch simulation, METROPOLIS requires the geometrical structures of the network with static congestion laws and the O-D matrix of all commuters as static simulation settings. By calibrating them, it is easy to introduce traffic events into the network, e.g. congestion of specific spatio-temporal structures.

### 3.2 Settings of IAURIF database

The network that we focus in IAURIF database covers totally 13627 links in Paris and its suburb region, as shown in Figure 2(a). There are totally 146 simulated traffic sequences in the data set. Each simulated traffic sequence covers 8 hours of traffic data observations, involving congestion in peak hour. Total 48 time sampling steps within each simulation divide the whole 8 hours into 15-minute bins over which the network traffic flows are aggregated. To represent local traffic states, we use traffic index [46][47] in Eq.8:

$$x_{pq} = \frac{\Delta t_p^0}{\Delta t_{pq}} \in [0,1] \qquad (8)$$

The denominator is the observed traveling time of link *p* at the time interval *q*. It is calculated by averaging all observed traveling time of commuters on the link *p* within the specific time interval. The numerator is the minimum traveling time among all commuters on this link within the given time interval. According to Eq.8, the traffic index belongs to [0,1]. As the value of the traffic index decreases, the corresponding link becomes more congested. We store the traffic index of each link at each time sampling step in matrix **X** containing 13627 rows and 7008 columns. Each column is the network-level traffic state observed at the same time interval, represented as a 13627 dimensional vector.

The simulated traffic sequences involve three different configurations of O-D matrix of all 3000 commuters in the network. They start their travel from outskirts of Paris into the central area within 8 hours in each simulation. With this setting, we aim to describe traffic behaviors during morning peak hour of the Paris transportation network. Different configurations of O-D matrix result in variations of global congestion level and different spatial distribution of congestion during peak hour. In the first setting, traffic demands are distributed relatively evenly in the outskirt area near the central Paris, leading to light isotropic congestion inside and around the central Paris. For the second case, we set more travel plans from the northern outskirt area to the central Paris, which produces local congestion patterns in both the northern outskirt and central region. Furthermore, we add random variance to the total amount of travel plans contained in the O-D matrix, covering both globally light and heavy congestion sharing the specific spatial congestion patterns in the network. In the third case, we increase travel paths inside the central and northern area to cause extremely heavy traffic burden in the corresponding areas. As a result, the derived traffic sequences suffer from global congestion ever since the beginning of simulation. They are used to simulate occurrence of extreme accidents in the network. We name the three settings as "Isotropic Traffic Demand" (ITD), "Anisotropic Traffic Demand" (ATD) and "Extreme Traffic Demand" (ETD). Following the three settings, we generate 37, 91, 18 simulations respectively.

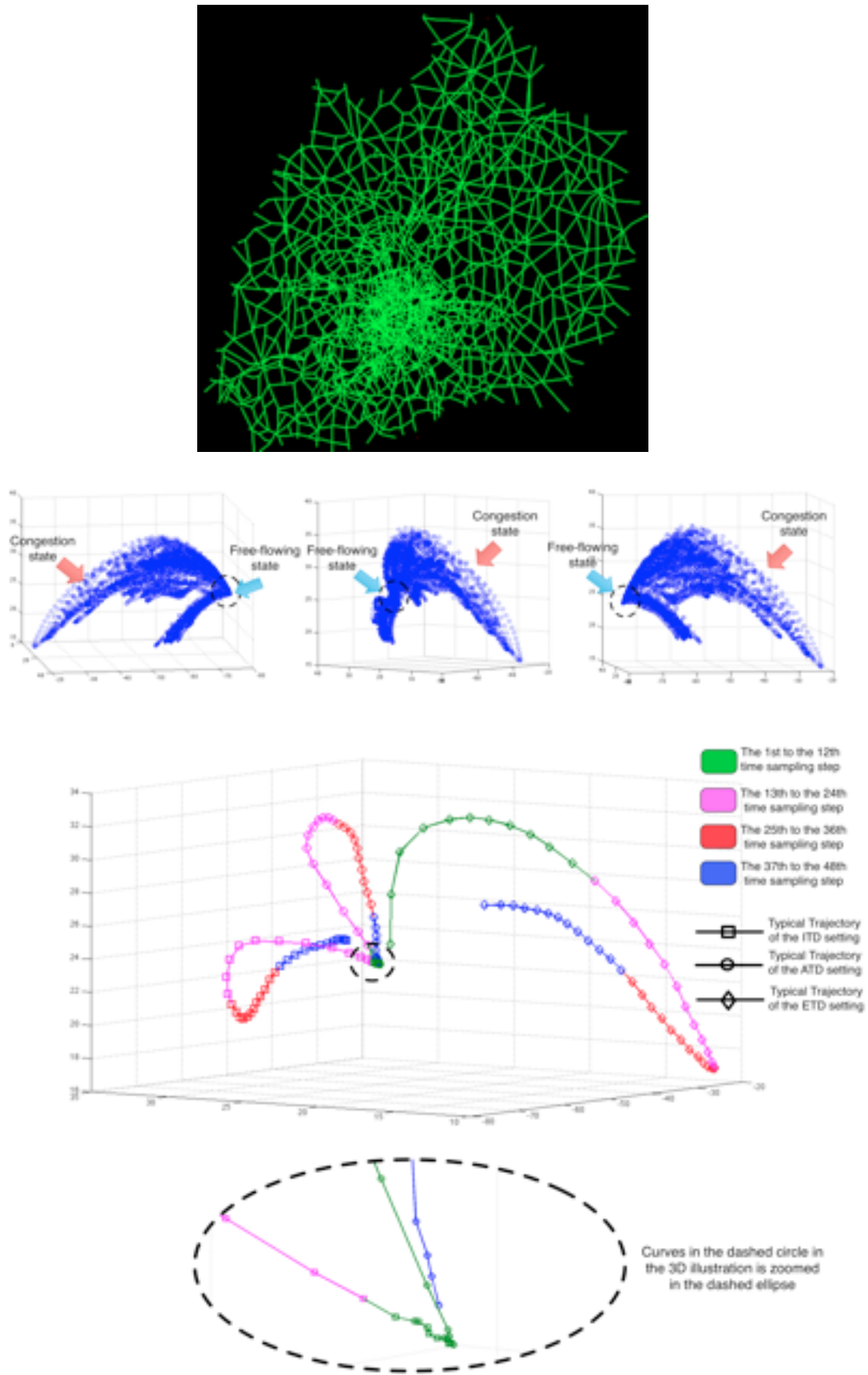

*Figure 2.* (a) Traffic network of Paris and suburb regions; (b) A three-views diagram of network-level traffic states in 3 dimensional PCA space; (c) Three typical trajectories

*corresponding to the three settings of traffic demands*

To visualize distribution of the network-level traffic states, we project all 13627 dimensional vectors into 3 dimensional PCA space, as shown in three different viewpoints in Figure 2(b). In our work, figures illustrating data distribution and clustering results in the PCA space, as Figure 2(b), 2(c), Figure 4, Figure 6 and Figure 8, the three axes correspond to the top three principal components that keep most variances of the original data. The observations of the global free flowing states are distributed within a small region compactly. In contrast, those of medium or severe congestion are distributed sparsely and biased from the region of the free flowing state. Spatial configurations of local traffic states keep the same if the whole network is global free flowing. On the contrary, congestion occurred at different parts of the network changes the spatial configurations in different ways, increasing variations of global traffic patterns. In Figure 2(c), we link network-level traffic states of the same simulated sequences following their temporal orders. The resultant trajectory represents temporal evolution of network-level traffic states. Different markers on trajectories are used to indicate different traffic demand settings. In each trajectory, color legends are used to indicate successive time intervals. The typical trajectories of the ITD and ATD setting have distinctively different orientations in the PCA space, consistent with difference of spatial congestion patterns. All trajectories start from global free flowing state, as we initialize all simulations with global free flowing state. Trajectories of the ITD and ATD settings converge to the free flowing state, indicating the network restores its fluidity after peak hour. In contrast, the ETD setting leads to much server congestion in the network, with some links congested even at the end of simulations. Thus the corresponding trajectory is quite different from the other two, and converges to the area located far from the free flowing state. Our clustering analysis involves two subjects. Firstly, we perform clustering on network-level traffic states, in order to unveil typical spatial configurations of network-level traffic states, as described in Section 4. After that, we adopt clustering on temporal trajectories of network-level traffic states in Section 5, based on which we could study large-scale traffic dynamics.

## 4. Spatial configurations of network-level traffic states

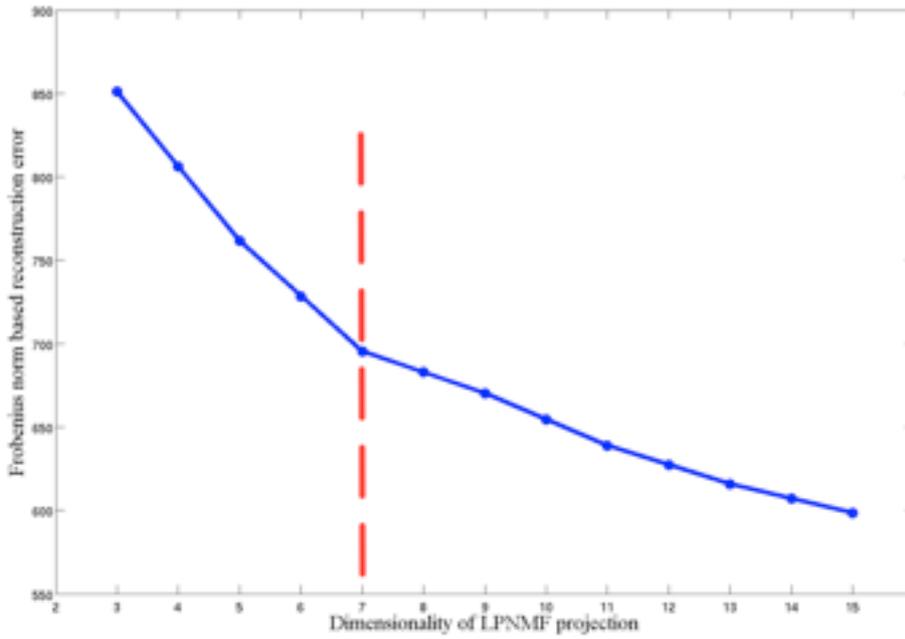

*Figure 3 Frobenius norm based reconstruction errors of different s*

In our work, each LPNMF projection is considered as *s* dimensional signature feature of global traffic configuration. To choose proper dimensionality of LPNMF projection for clustering, we evaluate the Frobenius norm based reconstruction error of the factorization results (shown in Eq.1) with different *s* ranging from 3 to 15, as shown in Figure 3. The reconstruction error declines much slower with *s* larger than 7, indicating 7 dimensional LPNMF projection is competent for describing network-level spatial congestion patterns. Therefore, we set *s* to be 7 in the followings.

In our clustering scheme, the number of the clusters *K* in K-means is decided by following a statistical compactness evaluation of the clusters. Given *p* derived clusters, the compactness $c$ of the clusters is evaluated using the average of sample variances of each cluster, as defined in Eq. 9:

$$c = \frac{1}{p}\sum_{i=1}^{p}(\frac{1}{N_i-1}\sum_{j=1}^{N_i}(nd_j - \frac{1}{N_i}\sum_{j=1}^{N_i}nd_j)) \qquad (9)$$

$N_i$ is the number of samples assigned into the *i*-th cluster. $\{nd_j\}$ are the network-level traffic observations of the *i*-th cluster. This criterion has been used in ward-linkage hierarchical clustering [49]. The average sample variance represents general level of compactness of clusters, which is used as the stopping criterion of hierarchical division of the cluster structure. The lower average sample variance is, the more compact clusters we obtain. In our case, we expect network-level traffic states sharing similar spatial configurations of congestion to present a compact cluster structure. To achieve this goal, for each *K* ranging from 3 to 7, we evaluate the compactness measure. Further larger *K* results in cluttered clustering structure, which is difficult to explain the correspondence between the derived clusters and underlying global traffic state patterns. According to Table 1, the compactness measure declines when *K* increases from 3 to 5. *K* larger than 5 introduces little change. It indicates that *K* equaling to 5 is a suitable setting to unveil typical network-level traffic state patterns.

*Table 1. Compactness measure of clustering results based on LPNMF and PCA*

| Algorithm | 3 | 4 | 5 | 6 | 7 |
|---|---|---|---|---|---|
| LPNMF | 182.0142 | 179.0684 | 171.7805 | 173.9921 | 175.6909 |
| PCA | 197.1148 | 182.7953 | 156.6382 | 154.2231 | 152.9623 |

Figure 4 illustrates the cluster structures in the PCA space. Figure 4(a) illustrates clustering results when *K equals to* 3. The cluster of green legends contains network-level traffic data between the 1$^{st}$ and 20$^{th}$ time sampling step of all 146 simulations, which counts 76% of all samples in the cluster. The left 24% of the cluster come from the interval ranging from the 35$^{th}$ to 48$^{th}$ time sampling step of 97 simulations. All data of the cluster are distributed within the region of the global free flowing state. We therefore name it as "Free Flowing Cluster" (FFC). The cluster of blue legends is composed mainly of traffic state observations collected between the $18^{th}$ to $40^{th}$ time sampling step from all 37 simulations of the ITD simulation setting. The cluster of the red legends consists of traffic data between the $20^{th}$ and $48^{th}$ time sampling step from the 105 simulations of the ATD and ETD setting. We thus name these two clusters as "Light Isotropic Congestion Cluster" (LICC) and "Anisotropic Congestion Cluster" (ACC) respectively. Both of them cover peak hour ranging from the $20^{th}$ until $40^{th}$ time sampling step. The corresponding three exemplars in the figure indicate typical spatial congestion patterns of the three clusters. In our work, we use the average spatial configuration of traffic indexes as the exemplar of the corresponding cluster, which is calculated by taking link-wise average over the 30% most congested network-level traffic state observations in each cluster. The congested links with traffic indexes lower than 0.79 are labeled using red legends, in order to make the spatial congestion pattern of each cluster visually distinctive. The LICC exemplar contains evenly distributed congestion around the outskirt and central region of Paris. In contrast, congestion concentrates in the central region and the northern outskirt in the ACC exemplar, which is consistent with characteristics of the ATD and ETD setting. In the figure, spatial traffic configurations of the ITD and ATD settings are separated perfectly in the clustering result. By increasing the number of clusters to 5 in Figure 4(b), we can find more details about spatial traffic configurations. The ACC cluster is split into three sub-clusters labeled by pink, purple and black legends respectively. Figure 5 illustrates exemplars of the obtained sub-clusters following the setting in Figure 4(a). Over 90% of the black-labeled cluster are collected between the $10^{th}$ and $48^{th}$ step of all 18 simulations of ETD setting. Due to the extremely heavy traffic demands of the ETD setting, congestion appears since the $10^{th}$ step until the end. The traffic configuration of this cluster presents severe congestion in the network. Consistently, the exemplar of this cluster shown in Figure 4(a) illustrated much severer congestion over the central region and its surroundings than the others. Therefore, we name it as "Heavy Congestion Cluster" (HCC). The pink-labeled cluster corresponds to the peak hour period ranging from the $15^{th}$ to $35^{th}$ time sampling step of 88 simulations following the ATD setting. In the purple-labeled cluster, data samples come from the time interval after peak congestion (from the $30^{th}$ time sampling step to the $48^{th}$ time sampling step) of 95 simulations. 75 of them are shared with the pink-labeled cluster. It means that the 75 sequences evolve from the pink-labeled cluster to the purple-labeled cluster successively during the peak hour period. Among the left 20 simulations, 3 of them correspond to the ATD setting but with heavier congestion. The other 17 simulations are derived based on the ETD setting and shared with the HCC cluster. The purple-labeled cluster covers the tails of the 17 simulated sequences after the peak of congestion, while the HCC cluster involves the peak hour interval. Therefore, we name the sub-clusters labeled by pink and purple legends as "Peak Congestion Cluster" (PCC) and "After-peak Congestion Cluster" (APCC) respectively. Figures 5(b) and 5(c) illustrate the exemplar of the two sub-clusters. The general congestion level of the PCC exemplar is heavier than the APCC exemplar. Both of them have similar congestion pattern inside the central Paris, while the PCC exemplar contains more congested links in the northern outskirt. The two sub-clusters indicate gradually restoration of traffic conditions during peak hour, representing a spatio-temporal traffic pattern in the network.

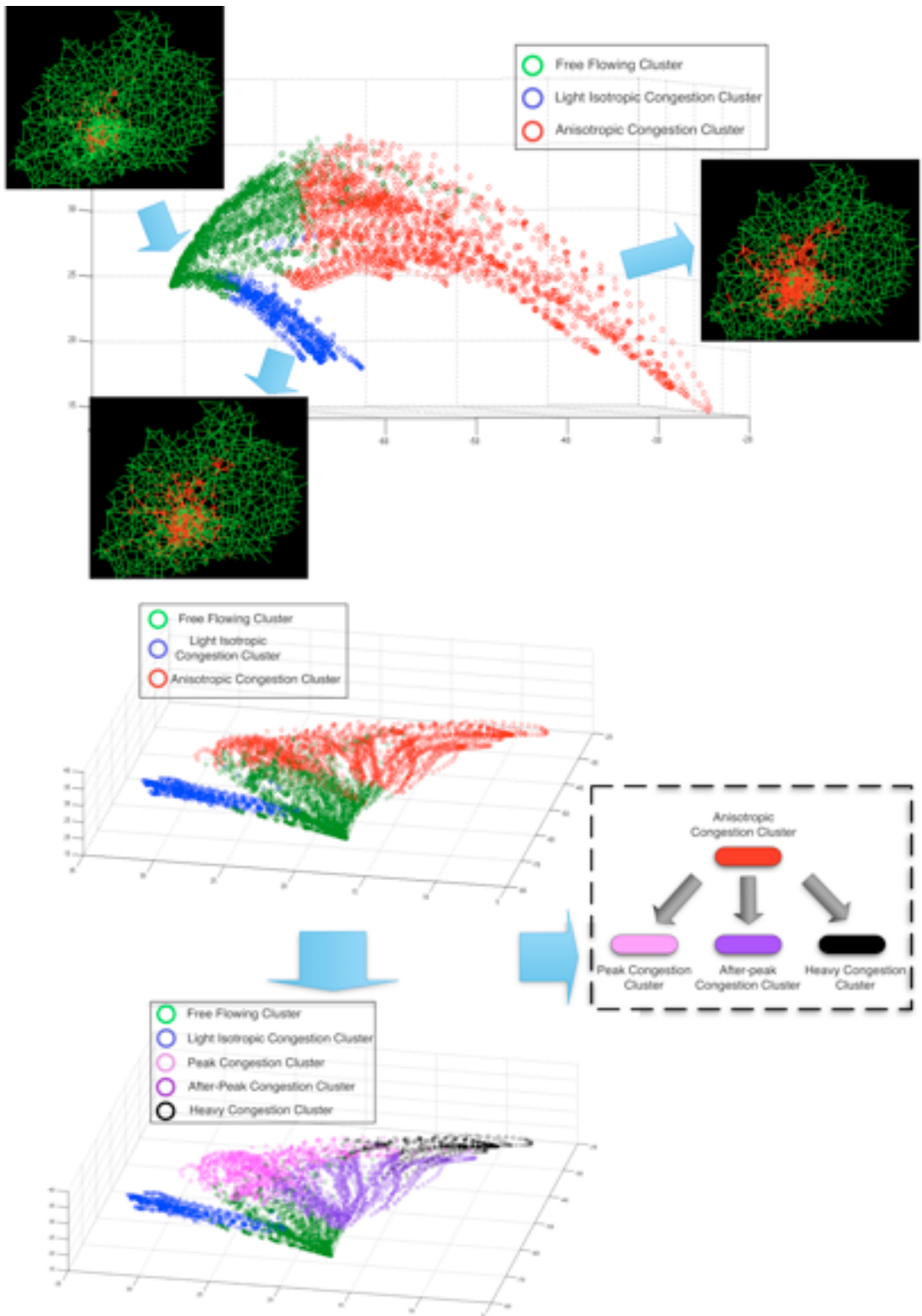

*Figure 4.* (a) Three clusters and exemplars of network-level traffic states; (b) Division of clusters after increasing the number of clusters to 5.

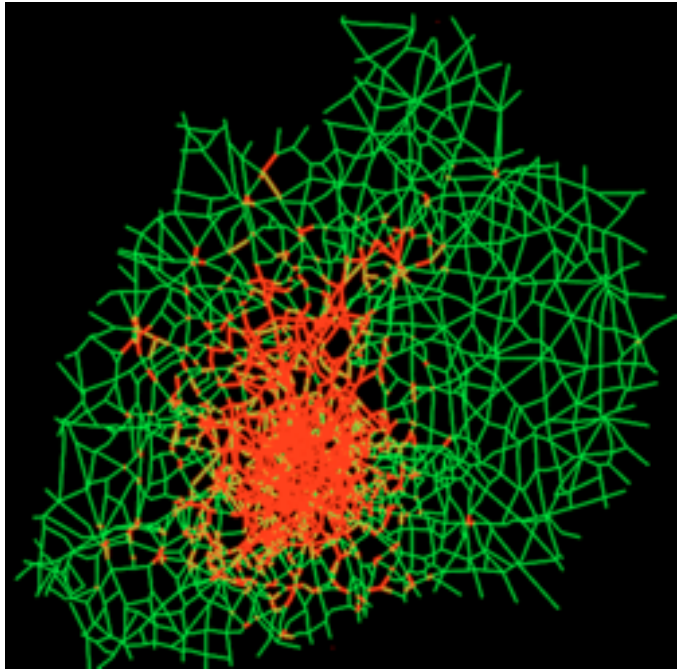

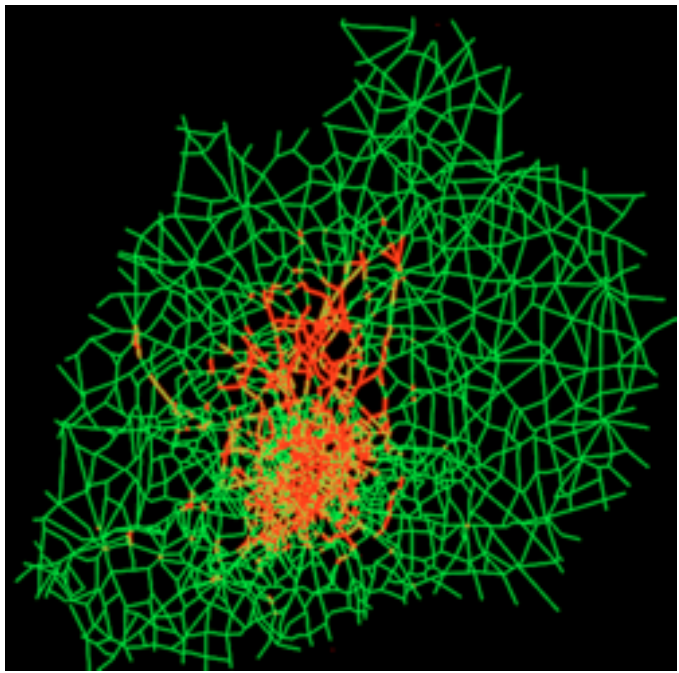

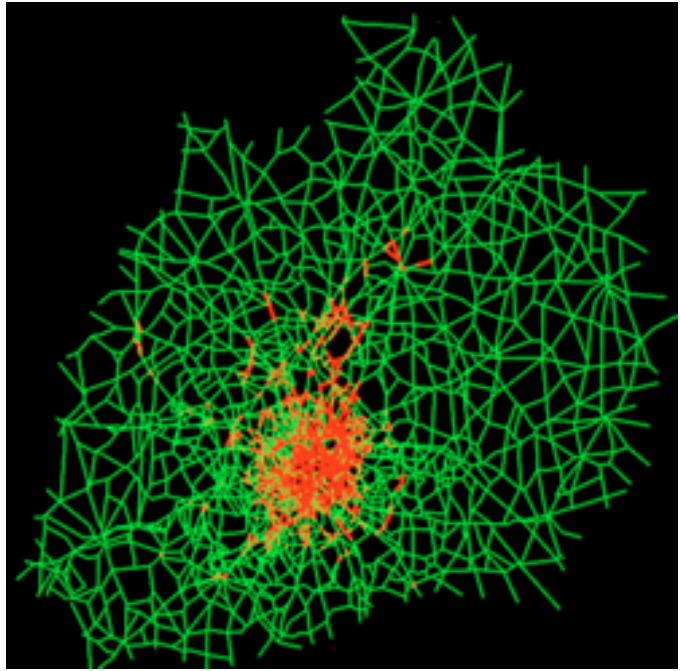

*Figure 5.* Exemplars of the sub-clusters: (a) the exemplar of the HCC cluster; (b) the exemplar of the PCC cluster; (c) the exemplar of the APCC cluster.

To verify capability of LPNMF in unveiling global traffic patterns, we perform K-means on *PCA projections* of the network-level traffic states and compare the derived clustering results. PCA is known as a baseline algorithm in manifold learning. For clustering, we keep the first 15 principal components that contain over 50% variance of the original data. The obtained 15 dimensional PCA projections are then used for clustering. For comparison, we vary *K* in K-means from 3 to 7. According to Table 1. PCA based clustering leads to higher average sample variances when *K* is 3 and 4. Figure 6 shows the derived clustering structures. As shown in the figure, with *K* to be 3, the PCA based clustering only indicates variations of average congestion level over the whole network, ignoring differences between spatial congestion patterns of the ITD and ATD settings. It results in high variances of the cluster structure. By increasing the number of clusters to 5, the PCA based clustering separates global traffic states of the two different simulation setting, thus obtains more compact clusters in statistical sense. However, the derived clusters of pink and purple legends in Figure 6 fail to identify the spatio-temporal structure of traffic patterns as shown in Figure 4(b). Compared with PCA, LPNMF is not only a dimension reduction tool, but also a feature extraction procedure to construct an informative representation of global traffic states.

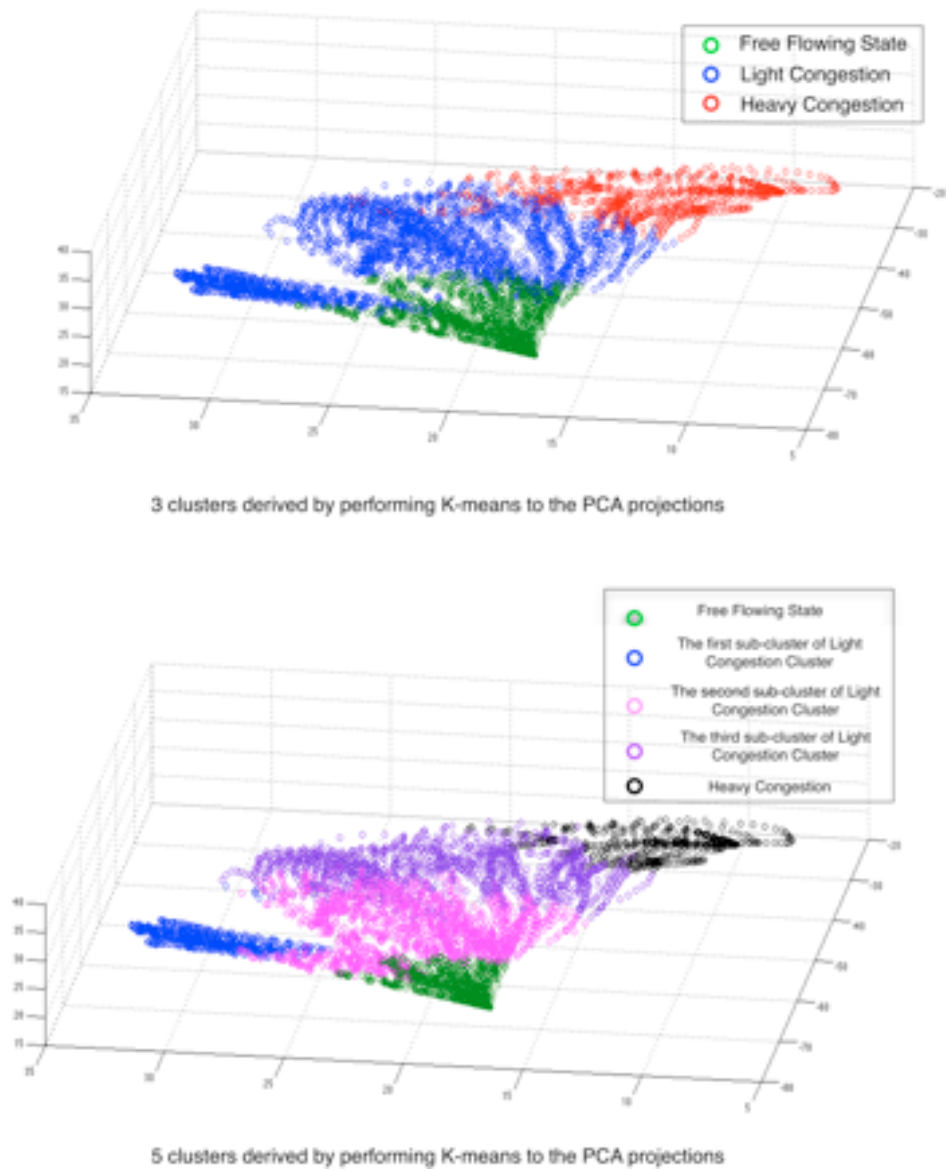

*Figure 6. Clustering results derived by performing K-means on PCA projections.*

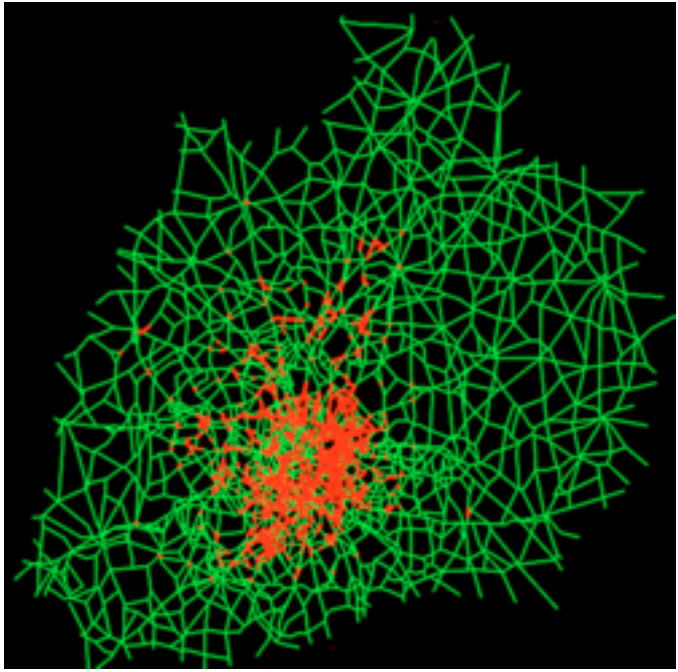
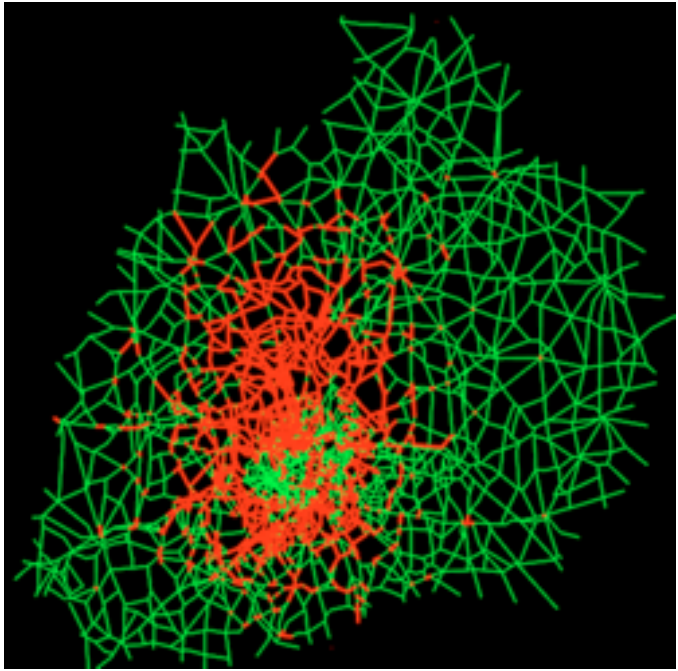

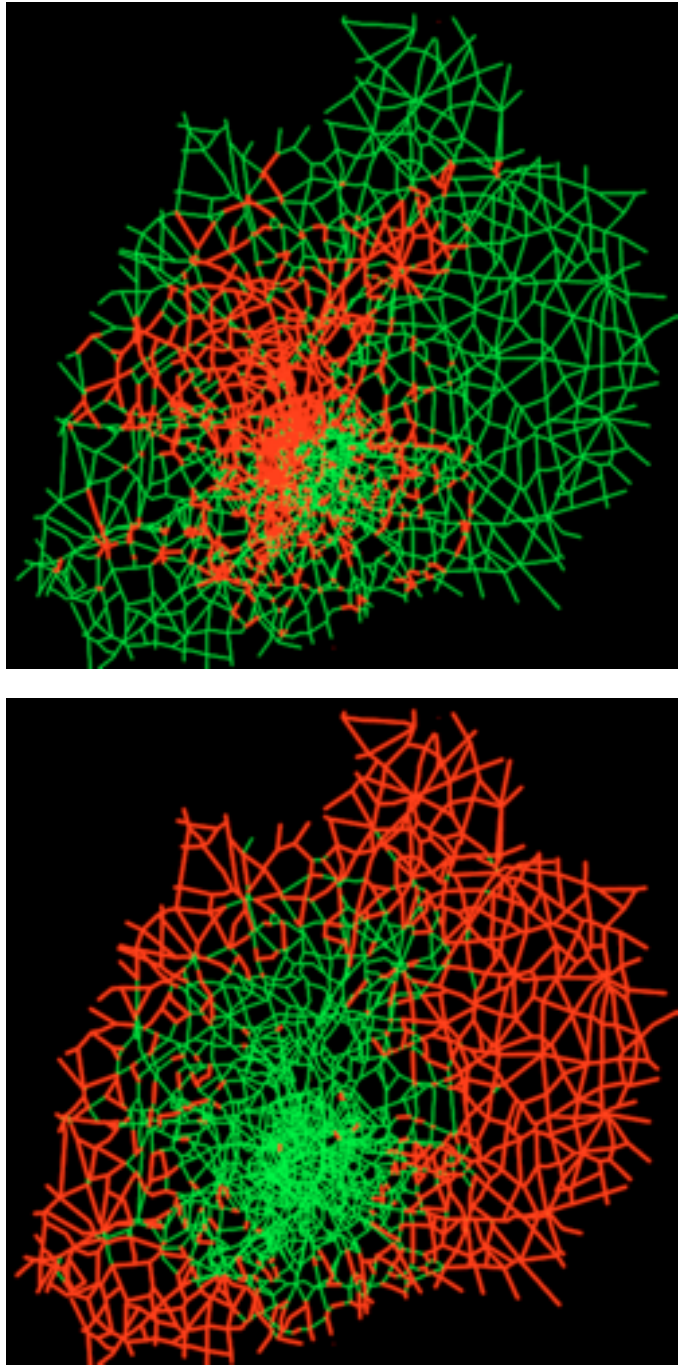

*Figure 7. (a) and (b) are LPNMF basis vectors corresponding to the two highest row-wise average values in **V**; (c) and (d) are LPNMF basis vectors corresponding to the two smallest row-wise average values in **V***

Besides clustering on the reduced LPNMF representation, we propose to investigate spatial layouts of LPNMF basis vectors that represent important components of global traffic configurations. We calculate the row-wise average $\frac{1}{m}\sum_{j=1}^{m} V_{i,j}$ and sort the *s* average values. The LPNMF basis vectors corresponding to the two largest and two smallest row-wise average values are selected. We analyze the physical significance of the selected basis vectors in the followings. The localized links with the top 20% largest entries (2725 entries) in each basis vector are labeled and their spatial locations are illustrated with red legends. We choose two LPNMF basis vectors corresponding to the two largest row-wise average in **V** and show their spatial layouts in Figures 7(a) and 7(b). Furthermore, Figures 7(c) and 7(d) show the spatial layouts of the basis vectors corresponding to the two lowest row-wise average values. The links with distinctively large LPNMF magnitudes correspond to the local regions that are highly correlated in forming the spatial traffic configurations. According to Section.2, the LPNMF basis vector with higher row-wise average value of **V** contributes more in representing the spatial congestion patterns. Following this idea, as shown in Figures 7(a) and 7(b), links in central Paris play the most critical role in constituting typical spatial distribution patterns of congestion. Compared with the central region, the outskirt area, especially the northern outskirt, has less but still important contribution in global traffic configurations. Since most travel paths are oriented to the central region in the simulations, Paris center plus its surroundings is expected to be the area that is more likely to be congested during peak hour. Therefore, congestion patterns in this region are the most important factors of global traffic configurations. On the contrary, the circular outskirt regions located far from the central area are free flowing at most times. They contribute the least in global traffic configurations, as shown consistently in Figures 7(c) and 7(d). The spatial layouts of the LPNMF basis vectors represent segmentation of the geographical structure of the network. Different separated regions have effects to different extents on yielding the global traffic state configurations. Links within the same region have homogeneous traffic characteristics. By looking into the basis vectors, we can identify traffic bottleneck of the whole network and extract spatial correlation patterns of the links. Such structural information provides a prior knowledge about how links are correlated with each other when we describe traffic dynamics of the network using graph models [50][51].

## 5. Temporal analysis of network-level traffic states

In this section, we analyze typical temporal dynamic patterns of network-level traffic states by performing clustering of large-scale traffic temporal behaviors in the 7 dimensional LPNMF projection space. This analysis is important in understanding how the global configuration of traffic states varies throughout a large time period. We also aim to investigate correspondences between the obtained typical temporal dynamic patterns and the underlined simulation settings, which further verifies ability of the LPNMF based method in analyzing global traffic dynamics.

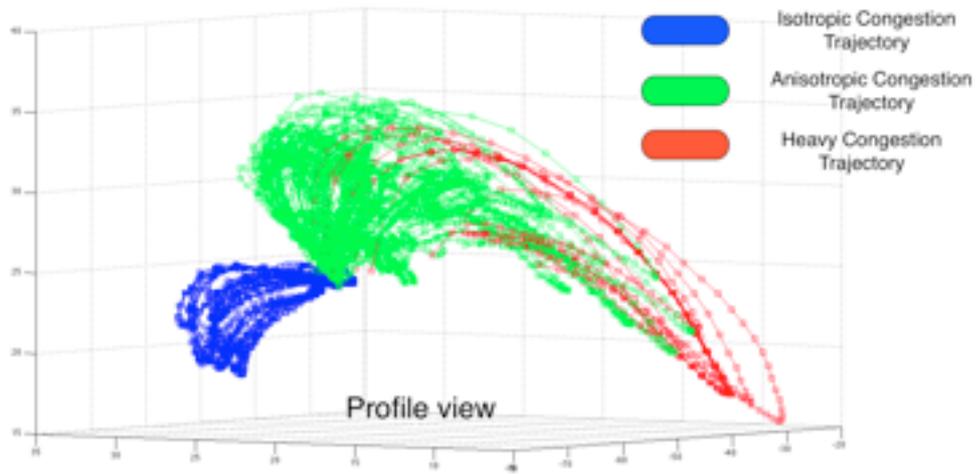
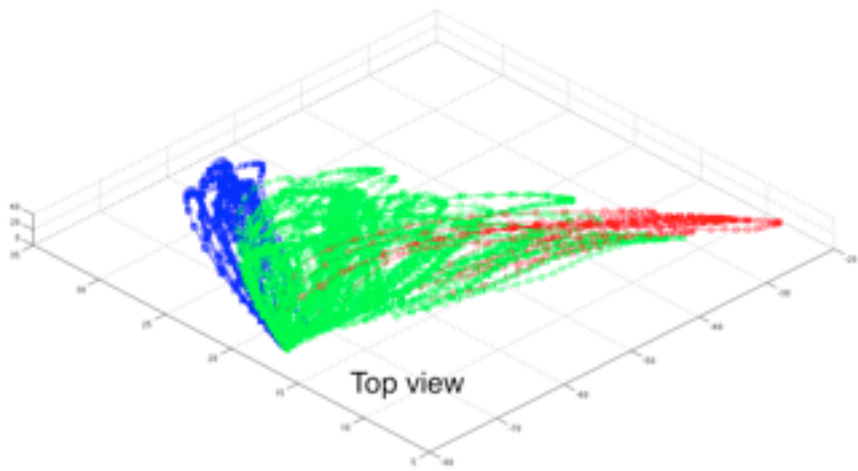

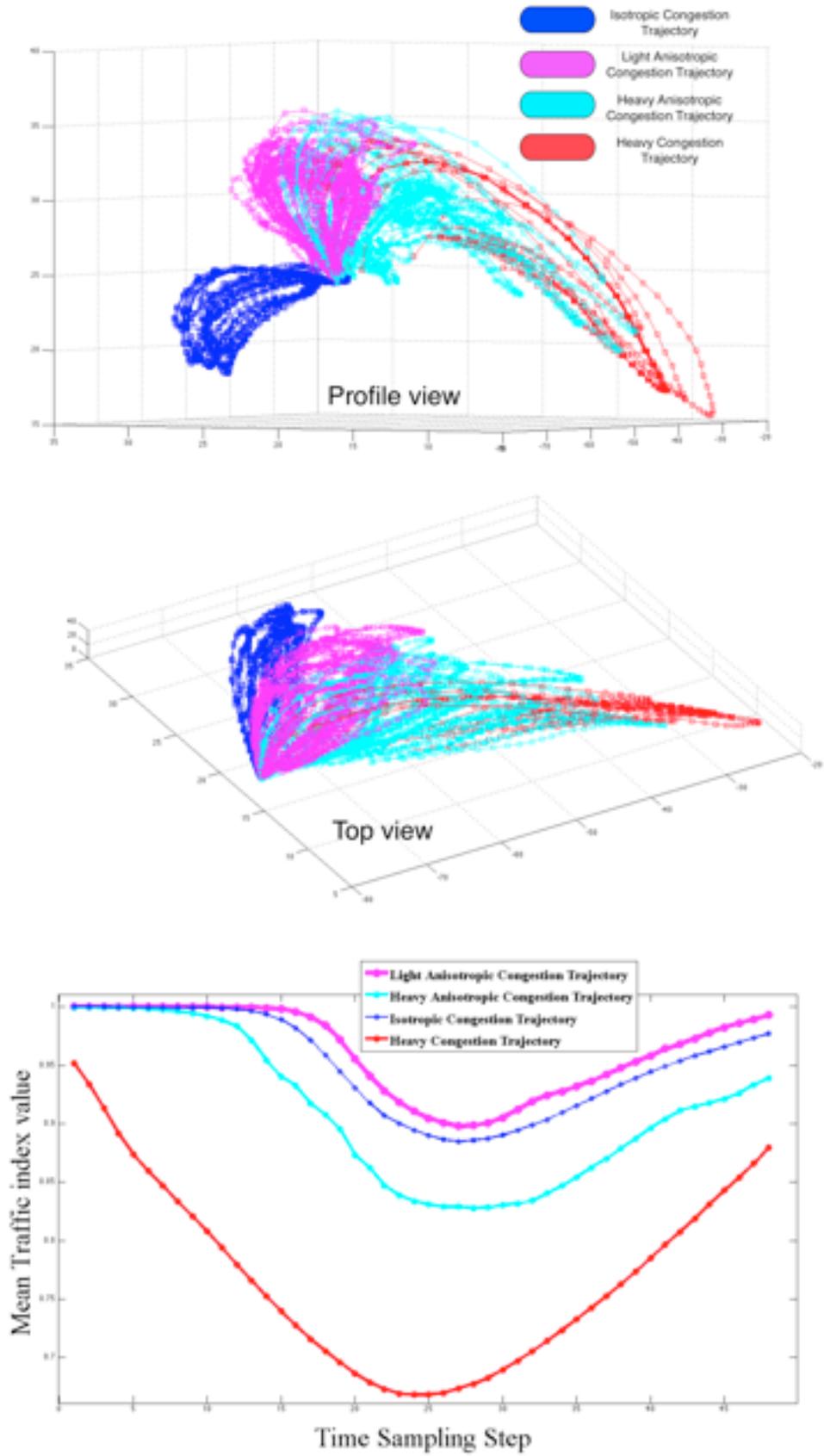

*Figure 8*. (a) Three clusters of the temporal trajectories in 3 dimensional PCA space; (b) Four clusters of the temporal trajectories; (c) Comparison of average temporal dynamic patterns of the four trajectory clusters.

For each studied traffic sequence, we represent the trajectory of network-level traffic states in the LPNMF space as $\{seq_i\}(i=1,2...48)$, with each $seq_i$ as a 7 dimensional LPNMF projection. To measure similarity between trajectories $\{seq_i^a\}$ and $\{seq_i^b\}$, we compute cosine distance [52] between the LPNMF projections of the corresponding time sampling steps and sum the distances derived in the whole time period, as defined in Eq.10:

$$D = \sum_{i=1}^{48} \frac{seq_i^a \cdot seq_i^b}{\|seq_i^a\|\|seq_i^b\|} \quad (10)$$

The cosine distance measures the cosine value of the angle between two corresponding LPNMF projection. It is well normalized into [0,1], which is easy to manipulate in computation. We perform K-means clustering on the temporal trajectories using the distance measure. We set *K* to be 3 firstly, in order to find correspondence between the trajectory clusters and the underlying simulation settings. Figure 8(a) illustrates the derived three clusters of the trajectories. The cluster labeled by blue legends consists of all 37 traffic sequences of the ITD setting, The black-labeled cluster consists of total 11 of 18 trajectories of the ETD setting. The left 7 of 18 ETD simulations are absorbed into the green-labeled cluster. 91 of total 98 trajectories in the green-labeled cluster correspond to the simulated sequences generated following the ATD setting. The obtained three trajectory clusters categorize trajectories of different traffic demand settings accurately. We name them henceforth by "Isotropic Congestion Trajectory" (ICT) and "Anisotropic Congestion Trajectory" (ACT) and "Heavy Congestion Trajectory" (HCT) respectively. By increasing the cluster number from 3 to 4, we can find the ACT cluster is divided further into two sub-clusters, corresponding to different general congestion level during peak hour, named as "Light Anisotropic Congestion Trajectory" (LACT) and "Heavy Anisotropic Congestion Trajectory" (HACT). Figure 8(b) shows the clustering results. For each time sampling step, we treat mean traffic index value (averaged over all 13627 links in the network) as a crude measure of global traffic state at the current time, the sequence of totally 48 mean index values in one simulation form a general evaluation of large-scale traffic dynamics of the simulation. We further take average of all the 48-D sequences of mean index values in each trajectory cluster. The resultant average sequence represents the general dynamic pattern of the corresponding cluster. Figure 8(c) illustrates average sequences of mean index values corresponding to the trajectory clusters. According to the figure, ICT, LACT and HACT have similar general traffic dynamics with differences in duration of congestion and peak congestion level. In HCT, the network starts to suffer from congestion since the beginning of simulations, which is much different from the others and consistent with the ETD simulation setting. Temporal clustering analysis provides a divide-and-conquer solution to describe underlined large-scale traffic dynamic patterns. Sequences in the same cluster share a common statistical dynamic characteristic of global traffic configurations. By extracting and modeling the typical dynamic process of each cluster using the feasible dynamical models, we are able to improve the controllability and observability of the large-scale traffic dynamic process.

## 6. Conclusions and perspectives

In this article, we propose and present a new traffic mining methodology for unveiling spatio-temporal traffic patterns, with large-scale modeling and long term forecasting as ultimate goals. Our experiments on large-scale simulated traffic data shows ability of our approach to unveil meaningful congestion patterns and typology of time evolutions; also illustrated is the clear advantage compared to using classical dimension reduction methods such as PCA.

In applications of traffic data analysis, there is still an open issue about developing the on-line NMF training scheme. Since traffic state observations arrive successively in the form of sequences. It is necessary to update the NMF based model after accumulating a certain number of traffic state observations, which makes the derived model consistent with time-varying traffic configurations of the network. More important from the application point of view, one of our main focuses for future work is exploiting LPNMF low-dimension representation for long-term traffic forecasting.

## Acknowledgement

This work was supported by the grant ANR-08-SYSC-017 from the French National Research Agency. The authors specially thank Cyril Furtlehner and Jean-Marc Lasgouttes for providing advice and the benchmark database used in this article.